\documentclass[aps,prb,reprint,superscriptaddress]{revtex4-2}
\usepackage{amsfonts}
\usepackage{amssymb}
\usepackage{amsmath}
\usepackage{graphicx}
\usepackage{dcolumn}
\usepackage{bm}
\usepackage{units}
\usepackage{multirow}
\usepackage{CJKutf8}
\usepackage{subfigure}
\usepackage{color}%
\usepackage{url}
\usepackage[colorlinks,linkcolor=red,anchorcolor=green,citecolor=blue]{hyperref}
\providecommand{\U}[1]{\protect\rule{.1in}{.1in}}
\usepackage{array}
\newcommand{\PreserveBackslash}[1]{\let\temp=\\#1\let\\=\temp}
\newcolumntype{C}[1]{>{\PreserveBackslash\centering}p{#1}}
\newcolumntype{R}[1]{>{\PreserveBackslash\raggedleft}p{#1}}
\newcolumntype{L}[1]{>{\PreserveBackslash\raggedright}p{#1}}
\begin{document}


\title{Giant octupole moment in magnetic multilayers}


\author{Chang Niu}
\affiliation{Key Laboratory of Quantum Materials and Devices of Ministry of Education, School of Physics, Southeast University, Nanjing 211189, China}

\author{Lulu Li}
\affiliation{Key Laboratory of Quantum Materials and Devices of Ministry of Education, School of Physics, Southeast University, Nanjing 211189, China}

\author{Yiqing Wang}
\affiliation{Key Laboratory of Quantum Materials and Devices of Ministry of Education, School of Physics, Southeast University, Nanjing 211189, China}

\author{Lei Wang}
\email{wanglei.icer@seu.edu.cn}
\affiliation{Key Laboratory of Quantum Materials and Devices of Ministry of Education, School of Physics, Southeast University, Nanjing 211189, China}

\author{Ke Xia}
\email{kexia@seu.edu.cn}
\affiliation{Key Laboratory of Quantum Materials and Devices of Ministry of Education, School of Physics, Southeast University, Nanjing 211189, China}


\begin{abstract}
Multipole moments serve as order parameters for characterizing higher-order magnetic effects in momentum space, providing a framework to describe diverse magnetic responses by extending the concept of magnetism. In this letter, we introduce a methodology to quantitatively determine the multipole moment contributions in anomalous Hall effect through angle-dependent anomalous Hall current, with explicit incorporation of discrete crystal symmetries. Our technique uniquely enables the investigation of octupole contribution in non-periodic systems, particularly at interfaces and surfaces. Typically, in (Ag$_{2}$Fe$_{5}$)$_{n}$ multilayers with quantum-well-engineered $k$-point selectivity, we observe an octupole-dominated anomalous Hall effect in conventional ferromagnetic materials, through first-principles calculations. These results fundamentally challenge the existing theoretical understanding of the anomalous Hall effect, showing that even the conventional contribution arises not only from the dipole moment (net magnetization). Furthermore, we establish practical control over the octupole contribution through two distinct approaches: interface engineering and magnetic ordering reconfiguration, opening new possibilities for manipulating higher-order transport effects.
\end{abstract}


\maketitle



%



The prediction of non-zero anomalous Hall effect (AHE) in non-collinear antiferromagnetic material Mn$_{3}$Ir had significantly expanded the research scope of the anomalous Hall effect~\cite{PhysRevLett.112.017205}. This breakthrough not only provided novel insights for designing next-generation antiferromagnetic spintronic devices, but also had sparked a global upsurge in research on quantum transport in antiferromagnets ~\cite{sivadas_gate-controllable_2016,zelezny_spin-polarized_2017,tsai_electrical_2020,smejkal_crystal_2020,bernevig_progress_2022,qin_room-temperature_2023,li_field-linear_2023,cao_-plane_2023,yoon_handedness_2023,liu_chern-insulator_2023,wernert_hall_2025}.

To systematically determine whether the AHE emerges in noncollinear antiferromagnets, Suzuki \textit{et al.} introduced the concept of cluster multipole (CMP) moment~\cite{PhysRevB.95.094406}. This framework generalizes the conventional dipole-based description (governed by net magnetization) by identifying higher-order multipole moments, as the key order parameters for AHE in noncollinear antiferromagnets, such as magnetic octupole moment. Notably, in altermagnets, spin-split band structures can be directly linked to magnetic multipole moments~\cite{PhysRevLett.132.176702}, and these moments can be reversibly switched via current-induced spin-orbit torque (SOT)~\cite{higo_perpendicular_2022}. Thus, multipole moments serve as effective magnetic moments in momentum space, providing a unified description of spin-dependent transport phenomena~\cite{yatsushiro_multipole_2021,yanagi_generation_2023,bhowal_ferroically_2024}.

In antiferromagnets, dipole contribution typically vanish due to the absence of net magnetization, allowing magnetic multipole moments to dominate in noncollinear antiferromagnets and altermagnets. Intriguingly, recent studies reveal that even collinear ferromagnets (e.g., Fe and Ni) exhibit an in-plane AHE mediated by higher-order multipole moments~\cite{liu2024multipolaranisotropyanomaloushall,peng2024observationinplaneanomaloushall}. However, the octupole contribution to the AHE in such systems remains modest—constituting only $\sim$10\% of the conventional AHE~\cite{liu2024multipolaranisotropyanomaloushall,peng2024observationinplaneanomaloushall}.

\begin{figure}[tp]
	\centering
	\includegraphics[width=0.5\textwidth]{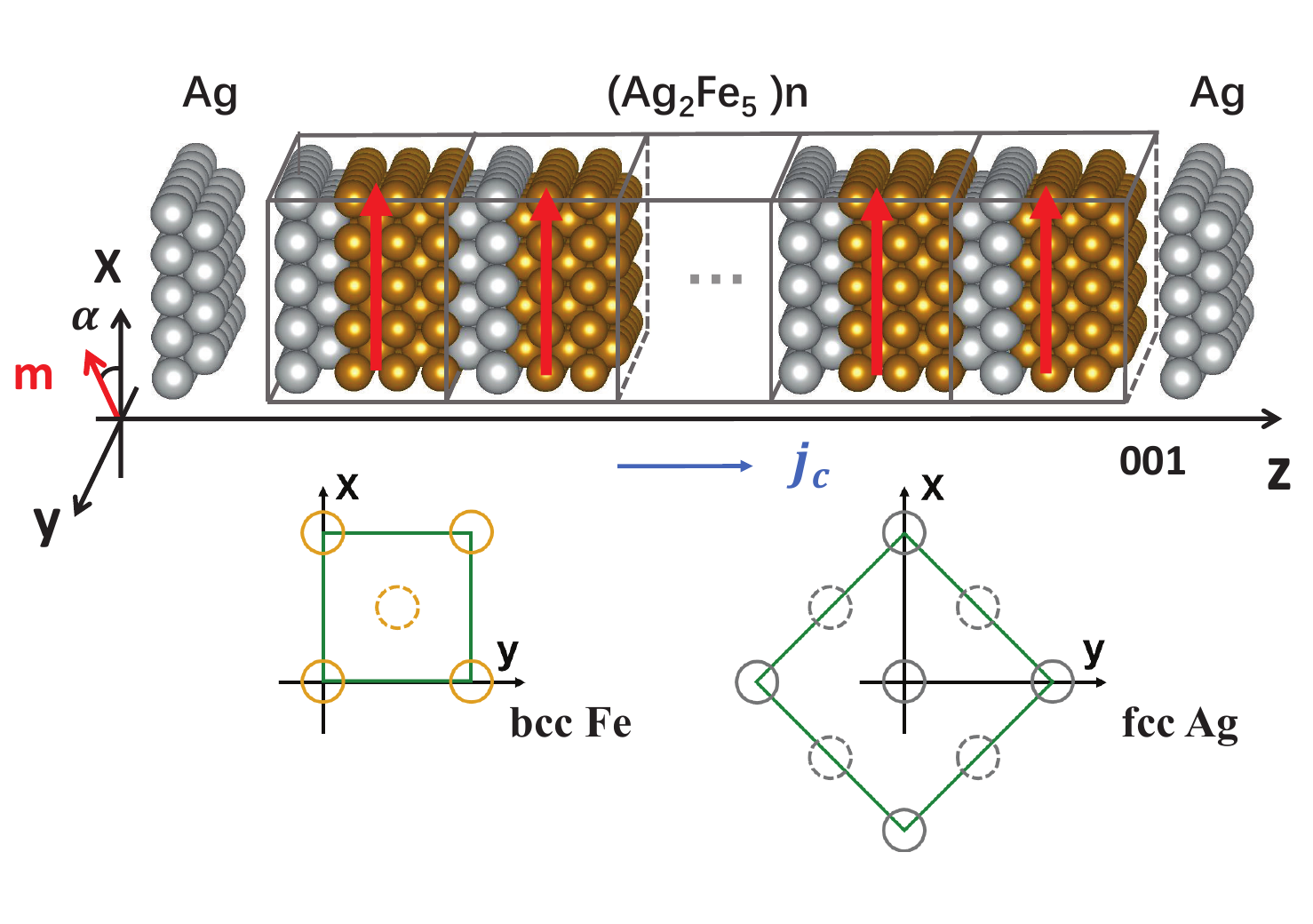} %
	\caption{The structure of (Ag$_{2}$Fe$_{5}$)$_{n}$ multilayer system along (001) direction, sandwiched by silver electrodes on both sides. The Ag lattice is rotated by 45$^\circ$ to match the Fe lattice. The injected current along the (001) direction (z-axis), the magnetization ($\mathbf{M}$) is rotated by an angle $\alpha$ in the ${xy}$-plane.}
	\label{fig1}
\end{figure}

\begin{figure}[htp]
	\centering
	\includegraphics[width=\columnwidth]{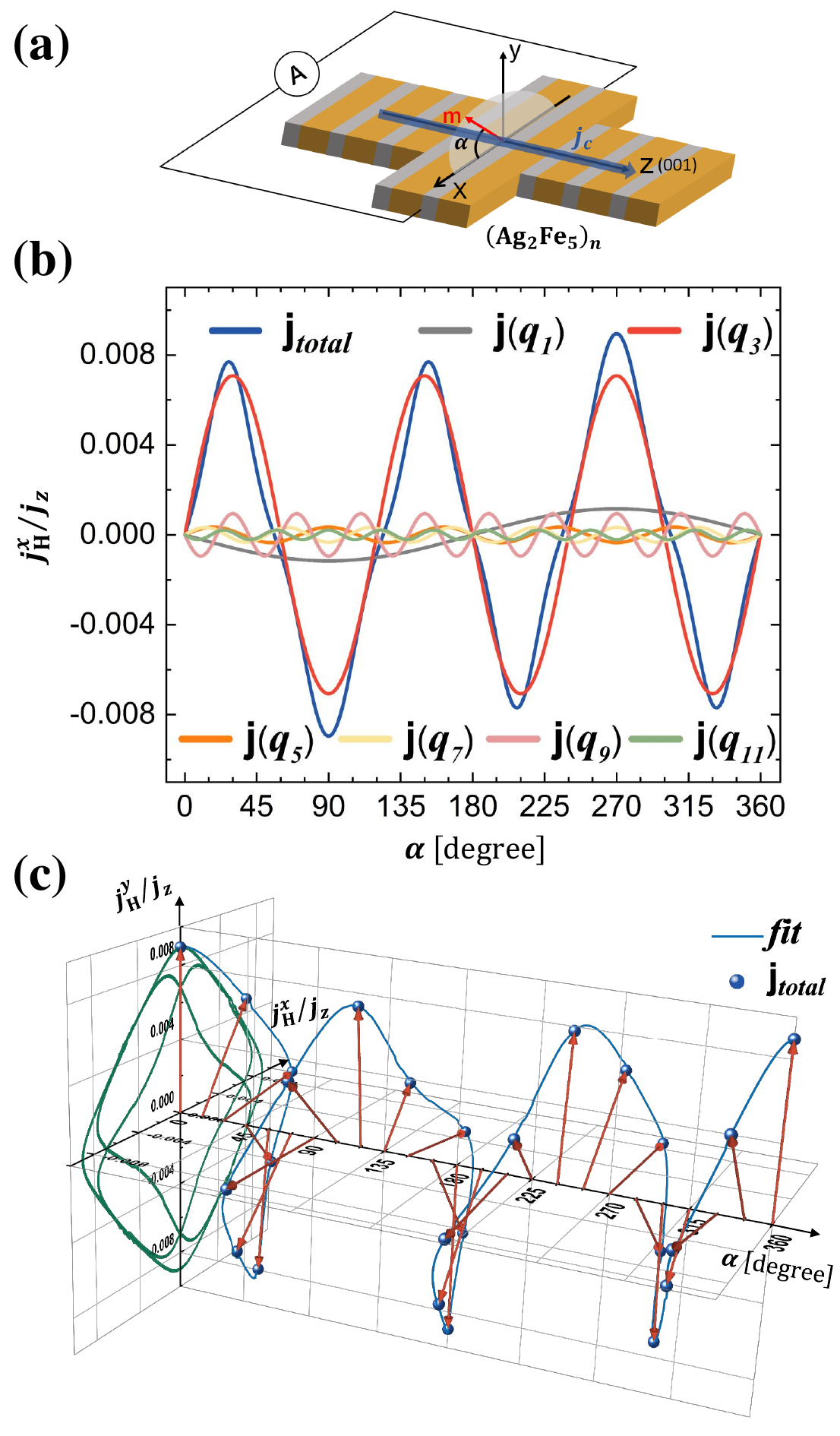} %
	\caption{(a) Hall measurement geometry for (Ag$_{2}$Fe$_{5}$)$_{n}$ multilayers with current injection along (001) direction ($z$-axis). Magnetization $\mathbf{M}$ rotates by angle $\alpha$ in the ${xy}$-plane. (b) Angular dependence of $j_H^x(\alpha)$ in (Ag$_{2}$Fe$_{5}$)$_{8}$, showing total current (blue) and $q_{i}$ contribution (colored), from Eq.~(\ref{eq4}). (c) Vector plot of $\bm{j}_H(\alpha)=(j_H^x,j_H^y)$ with current directions indicated (red arrows).
	}
	\label{fig2}
\end{figure}

Recent investigations of the AHE incorporating discrete crystal symmetry~\cite{li2024angulardependenceanomaloushall} have established a real-space relationship between the anomalous Hall current and magnetization direction, in contrast to the momentum-space description of the CMP theory. This approach enables studies of non-periodic systems, including interfaces and surfaces. Within this framework, the anomalous Hall current can be expressed as a higher-order multiple-angle function of the magnetization direction through tensor theory combined with discrete symmetry analysis. Given that both the tensor theory~\cite{li2024angulardependenceanomaloushall} and the multipole moment theory~\cite{liu2024multipolaranisotropyanomaloushall,peng2024observationinplaneanomaloushall,xiao2025anomaloushallneeltexturesaltermagnetic} exhibit angular dependence, we suggest a possible fundamental connection between these approaches.

In this letter, for a specific system with in-plane quadruple rotational symmetry ($C_{4v}$), we establish a connection between the angular-dependent linear response coefficients with multipole moments. Using this theoretical framework, we conduct first-principles calculations on (Ag$_{2}$Fe$_{5}$)$_{n}$ multilayer systems, which was engineered to host quantum well for $k$-point selecting. Our results demonstrate that the AHE is predominantly governed by octupole contribution, characterized by complex higher-order angular dependence patterns. Moreover, the octupole moment can be effectively modulated through interface engineering and magnetic ordering reconfiguration.

\begin{figure*}[htp]
	\includegraphics[width=1.6\columnwidth]{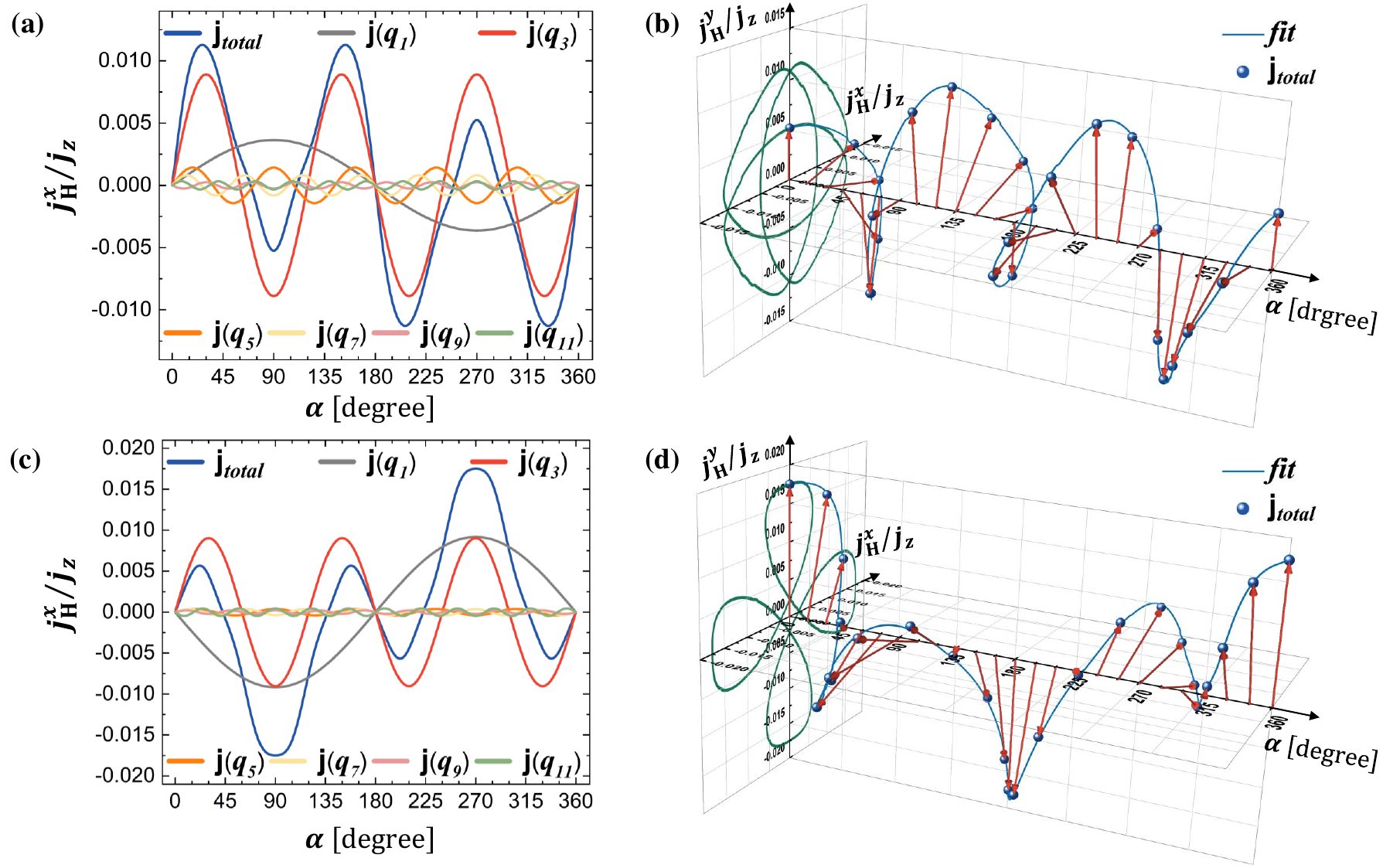}
	\caption{
	Angular dependence of anomalous Hall current in (Ag$_{2}$Fe$_{5}$)$_{n}$ multilayers. (a,c) Anomalous Hall current $j_H^x$ versus magnetization angle $\alpha$ for $n=6,10$ periods, showing total current (blue) and contributions from different-order coefficients $q_{i}$ (colored lines). (b,d) Vector plots of $\bm{j}_H(\alpha)=(j_H^x,j_H^y)$ derived from Eq.~(\ref{eq4}), with red arrows indicating current direction for selected $\alpha$ values.
	}
	\label{fig3}
\end{figure*}


Within the framework of tensor theory incorporating discrete crystal symmetry~\cite{li2024angulardependenceanomaloushall,rug01:000711810,1058782,WangXR2023,10.1063/5.0187589}, the linear response relation between the longitudinal and transverse charge current can be expressed as $\mathbf{j}_H=\hat{\Theta}\mathbf{j}_c$. Here, the linear response coefficients $\hat{\Theta}$ (effective Hall angle matrix) can be expanded in terms of the magnetization $\mathbf{M}$ as follows:
\begin{eqnarray}
	\hat{\Theta}_{ij}(\mathbf{M})=\hat{\Theta}_{ij}^0+\hat{\Theta}_{ij}^{k}\hat{M}_k+\hat{\Theta}_{ij}^{kl}\hat{M}_k\hat{M}_l+\cdots
	\label{eq.theta}
\end{eqnarray}
where $\hat{\Theta}_{ij}^{k}=\partial\hat{\Theta}_{ij}/\partial \hat{M}_k$ with $i,j,k\in\{x,y,z\}$, and the Einstein summation convention over repeated indices is employed. Notably, the above expression corresponds to a conventional Taylor expansion in the magnetization, where the physical meaning of the higher-order coefficients $\hat{\Theta}_{ij}^{kl\cdots}$ is ambiguous.

Recently, along with the rising of the multipole moment concept, the anomalous Hall conductivity vector, $\bm{\sigma}^H=(\bm{\sigma}^H_x,\bm{\sigma}^H_y,\bm{\sigma}^H_z)$, can be systematically formulated via multipolar expansion in the magnetic-order space~\cite{liu2024multipolaranisotropyanomaloushall,peng2024observationinplaneanomaloushall,xiao2025anomaloushallneeltexturesaltermagnetic}:
\begin{eqnarray}
	\begin{split}
	\bm{\sigma}_{i}^{H}=p_{ij}\hat{M} _{j} + o_{ijkl}\hat{M} _{j}\hat{M} _{k}\hat{M} _{l} +\cdots
	\end{split}
	\label{eq.sigma}
\end{eqnarray}
where $i,j,k,l\in\{x,y,z\}$, the dipole term $p_{ij}$ transforms as a rank-2 tensor, describing the linear response of the component of $\bm{\sigma}_{i}  ^{H}$ to the magnetization direction $\hat{M} _{j}$ along $j$-axis, while the octupole term $o_{ijkl}$ transforms as rank-4 tensor characterizing the higher-order response of $\bm{\sigma}_{i}  ^{H}$ to the magnetization direction $\hat{M} _{j}\hat{M} _{k}\hat{M} _{l}$, which satisfies symmetric and traceless with respect to the last three indices $j,k,l$. In this sense, the anomalous Hall current is expressed as $\mathbf{j}_H=\bm{\sigma}^H\times\mathbf{E}$, where $\mathbf{E}$ represents the applied electric field. Typically, when $\mathbf{j}_c\parallel\mathbf{E}\parallel z$, we have $\bm{\sigma}_{y}^{H}\equiv\hat{\Theta}_{xz}$ and $\bm{\sigma}_{x}^{H}\equiv-\hat{\Theta}_{yz}$. By comparing Eq.~(\ref{eq.theta}) and (\ref{eq.sigma}), we establish the correspondences  $p_{Yk}\equiv\hat{\Theta}_{xz}^k$, $o_{Yklm}=\hat{\Theta}_{xz}^{klm}$, $p_{Xk}\equiv-\hat{\Theta}_{yz}^k$, and $o_{Xklm}=-\hat{\Theta}_{yz}^{klm}$, etc. Therefore, the above coefficients in the tensor theory explicitly come from the multipole contributions and the other terms will be vanished due to the constrains from the crystal symmetry~\cite{li2024angulardependenceanomaloushall}.

For structures with in-plane quadruple rotational symmetry ($C_{4v}$) , as shown in Fig.~\ref{fig1}, both the above expressions of $\mathbf{j}_H=(j_H^x,j_H^y)$ can be expressed in terms of the magnetization direction $\mathbf{M}=(\cos\alpha,\sin\alpha,0)$~\cite{li2024angulardependenceanomaloushall,liu2024multipolaranisotropyanomaloushall,peng2024observationinplaneanomaloushall,xiao2025anomaloushallneeltexturesaltermagnetic}:
\begin{equation}
	\begin{split}
		\mathbf{j} _{H}/j_{z}&=\begin{bmatrix}
			j_{H}^{x}  /j_{z} \\
			j_{H}^{y}  /j_{z}
		\end{bmatrix}\\
		&=\sum_{i=4n-3}^{n\in [1,\infty )}q_{i}\begin{bmatrix}
			\sin i\alpha \\-\cos i\alpha
		\end{bmatrix} + \sum_{j=4n-1}^{n\in [1,\infty )}q_{j}\begin{bmatrix}
			\sin j\alpha \\\cos j\alpha
		\end{bmatrix}
	\end{split}
	\label{eq4}
\end{equation}
where $j_z$ is the injected charge current along z-axis, $\alpha$ is the angle between magnetization direction $\mathbf{M}$ and $x$-axis, and the first two coefficients are given by:
\begin{align}
	q_{1}&= \frac{1}{4} \left (4p_{Xx}+3o_{Xxxx}+3o_{Xxyy} \right )
	\\
	q_{3}&= \frac{1}{4} \left (-o_{Xxxx}+3o_{Xxyy} \right ).
\end{align}
Here, the dipole component $p_{Xx}$ describes the linear relationship between the net magnetization and AHE conductivity, while $o_{Xxxx}$ and $o_{Xxyy}$ characterize third-order octupole contribution, respectively.

Notably, the first-order coefficient $q_{1}$ corresponds to the conventionl AHE (consistent with $\mathbf{j} _{H}=q_{1} \mathbf{M} \times \mathbf{j} _{c}$), which is associated with both dipole and octupole moment. Considering the conventional AHE in ferromagnetic materials is generally treated as a response of the magnetization (dipole moment), our results indicate that even the conventional theoretical description of AHE requires revision. Moreover, the higher-order coefficients $q_{i}$ $(i\ne1)$, notably the third-order coefficient $q_{3}$ exclusively derived from octupole contribution. Therefore, we can study the contribution of octupole moment to the AHE by investigating the angular dependence of the anomalous Hall current in real space, which is applicable in non-periodic interfaces and surfaces.

Specifically, we designed (Ag$_{2}$Fe$_{5}$)$_{n}$ multilayer systems with alternating body-centered cubic (bcc) Fe and face-centered cubic (fcc) Ag layers, epitaxially stacked between semi-infinite Ag electrodes, as illustrated in Fig.~\ref{fig1}. To achieve lattice matching, the Ag lattice was rotated by 45$^\circ$ relative to the Fe lattice, preserving in-plane quadruple rotational symmetry ($C_{4v}$). The corresponding transport properties are determined from first-principles calculations via the fully relativistic exact muffin-tin orbitals (FR-EMTO) approach~\cite{wang_first-principles_2021,wang_abnormal_2023,wang_crystal-induced_2022}, implemented within the framework of scattering wave functions~\cite{starikov_calculating_2018,PhysRevB.99.144409,xia_first-principles_2006,ando_quantum_1991}.

We investigated the current density distributions in both longitudinal and transverse current by calculating the electron transport between any two atoms in the system. The interatomic charge transport processes can be quantitatively described by the following formula~\cite{wang_first-principles_2021,wang_crystal-induced_2022,wang_abnormal_2023,PhysRevB.99.144409,PhysRevLett.116.196602}:
\begin{eqnarray}
	\begin{aligned}
		J_{RR'}=\frac{1}{i\hbar}[\langle\Psi_{R}|\hat{\mathcal H}_{RR^{\prime}}|\Psi_{R^{\prime}}\rangle-\langle\Psi_{R^{\prime}}|\hat{\mathcal H}_{R^{\prime}R}|\Psi_{R}\rangle]
	\end{aligned}
\end{eqnarray}
where \( R \) and \( R^{\prime} \) denote two different atoms, \( | \Psi_R \rangle \) represents the scattering wave functions \cite{starikov_calculating_2018,PhysRevB.99.144409,xia_first-principles_2006,ando_quantum_1991} localized at site \( R \), and \( \hat{\mathcal{H}}_{R^{\prime}R} \) corresponds to the hopping Hamiltonian between atoms \( R \) and \( R^{\prime} \). By projecting the charge current \( J_{RR^{\prime}} \) onto different directions, we obtained the primary charge current density \( \mathbf{j}_c \) and the anomalous Hall current \( \mathbf{j}_H \). The anomalous Hall angle \( \Theta \) was then directly calculated as the ratio \( \Theta = \mathbf{j}_H / \mathbf{j}_c \). In detail, we calculate the components of the anomalous Hall current $\mathbf{j}_H=(j_H^x,j_H^y)$ by rotating the magnetization vector $\mathbf{M}$ within the $xy$-plane. Following systematic convergence tests, 5120×5120 $k_{\parallel }$ points in the lateral Brillouin zone (BZ) was employed to ensure a balance between the precision of Brillouin zone integration and computational efficiency.

Fig.~\ref{fig2}(a) schematically depicts the experimental configuration for measuring the anomalous Hall current $j_H^x$ in (Ag$_{2}$Fe$_{5}$)$_{n}$ multilayer systems. In this setup, the charge current $j_{c}$ is injected along the (001) direction ($z$-axis), while the magnetization $\mathbf{M}$ rotates in the $xy$-plane with an angle $\alpha$ relative to the $x$-axis. Using Eq.~(\ref{eq4}), we analyzed both the total anomalous Hall current and the individual contributions from different-order coefficients $q_{i}$ to $j_H^x$ in the (Ag$_{2}$Fe$_{5}$)$_{8}$ multilayer system, as presented in Fig.~\ref{fig2}(b). The blue curve shows the total anomalous Hall current, while the other colored curves represent the contributions from various order coefficients: for instance, the gray curve corresponds to the first-order coefficient $q_{1}$, and the red curve to the third-order coefficient $q_{3}$. Remarkably, the third-order contribution $q_{3}$ - associated with the octupole moment - accounts for nearly the entire anomalous Hall current, whereas other contributions are negligible. This striking observation reveals that the anomalous Hall effect (AHE) in (Ag$_{2}$Fe$_{5}$)$_{8}$ multilayers is dominated by the octupole moment, in contrast to conventional dipole-dominated ferromagnetic systems.

To better illustrate the dependence of the calculated anomalous Hall current $\mathbf{j}_H=(j_H^x,j_H^y)$ on the magnetization direction, Fig.~\ref{fig2} (c) presents the corresponding angular patterns: blue spheres represent the calculated total anomalous Hall current, and the blue line corresponds to the fitting results derived from the Eq.~(\ref{eq4}), the red arrows indicate the direction of the anomalous Hall current under different magnetization orientations ($\alpha$). It is evident that as $\mathbf{M}$ rotates in the $xy$-plane ($\alpha\in[0,360^\circ]$), the components of the anomalous Hall current, $\mathbf{j}_H=(j_H^x,j_H^y)$, exhibit clear $C_{4v}$ symmetry. In contrast to the nearly circular pattern expected from the first-order coefficient $q_{1}$, the outstanding contribution of octupole moment induces a complex nested structure in the fitted pattern, demonstrating pronounced higher-order harmonic angular dependencies.

To investigate the influence of periodic repeating unit number on octupole moment in (Ag$_{2}$Fe$_{5}$)$_{n}$ multilayer systems, we constructed two additional structural configurations with $n=6$ and $n=10$. Their angular dependence characteristics are shown in Fig.~\ref{fig3}(a)(b) and Fig.~\ref{fig3}(c)(d), respectively. Compared to the octupole contribution to the anomalous Hall current in (Ag$_{2}$Fe$_{5}$)$_{8}$, that in (Ag$_{2}$Fe$_{5}$)$_{6}$ is significantly reduced—yet still more than twice as large as the conventional AHE coefficient ($q_1$), as shown in Fig.~\ref{fig3}(a). Furthermore, Fig.~\ref{fig3}(c) reveals that the octupole contribution in (Ag$_{2}$Fe$_{5}$)$_{10}$ is even smaller, only slightly exceeding that of the conventional AHE coefficient ($q_1$). As illustrated in Fig.~\ref{fig3}(b)(d), the fitting results based on Eq.~(\ref{eq4}) demonstrate excellent consistency with first-principles calculations across all periodic configurations. Despite their distinct appearances, these angle-dependent patterns exhibit complex nested structures, highlighting the dominant contribution of octupole moment, while all preserving quadruple rotational symmetry ($C_{4v}$).
\begin{figure}
	\includegraphics[width=0.48\textwidth]{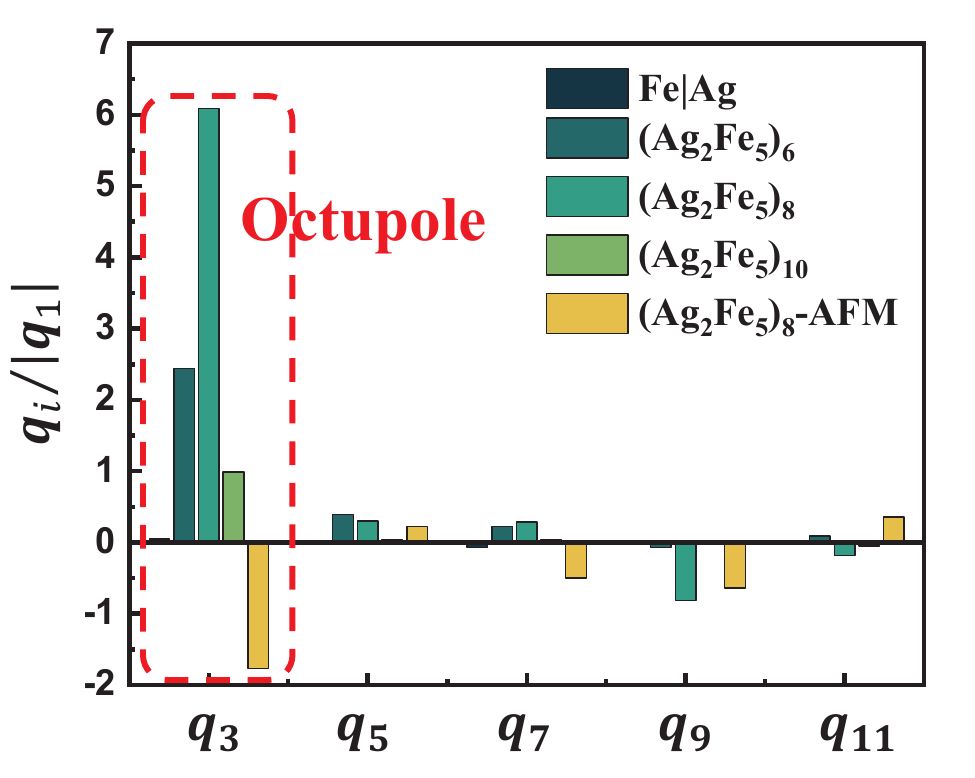}
	\caption{The fitting parameters normalized by the absolute value of $q_{1}$ for different systems along (001) direction using Eq.~(\ref{eq4}). Higher-order coefficients $q_{i}$ $(i\ne1)$ represent different multipole moments, notably the third-order coefficient \(q_{3}\) exclusively derived from octupole contribution. Typically, in (Ag$_{2}$Fe$_{5}$)$_{8}$ multilayer system, we also revere its ferromagnetic ordering of adjacent repeating units to construct an antiferromagnetic ordering, labeled as ``(Ag$_{2}$Fe$_{5}$)$_{8}$-AFM''.}
	\label{fig4}
\end{figure}

\begin{table}
	\centering
	\setlength{\tabcolsep}{28pt}
	\begin{tabular}{l r} 
		\hline
		\hline
		\multicolumn{1}{c}{} & \multicolumn{1}{c}{$q_{1}$} \\ 
		\hline
		Fe$\mid$Ag & 0.0578 \\
		\hline
		(Ag$_{2}$Fe$_{5}$)$_{6}$  & 0.0036 \\
		\hline
		(Ag$_{2}$Fe$_{5}$)$_{8}$  & -0.0012 \\
		\hline
		(Ag$_{2}$Fe$_{5}$)$_{10}$ & -0.0092 \\
		\hline
		(Ag$_{2}$Fe$_{5}$)$_{8}$-AFM & -0.0081 \\
		\hline
		\hline
	\end{tabular}
	\caption{The leading-order fitting parameter $q_{1}$ for different systems in Fig.~\ref{fig4}.}
	\label{tab1}
\end{table}

To elucidate the modulatory effects of interface engineering on the octupole moment, we present the normalized high-order fitting parameters in the bar chart, as shown in Fig.~\ref{fig4}. Here, normalization with the absolute value of $q_{1}$ guarantees the accuracy of the directions of the fitting parameters, and the values of the fitting parameter $q_{1}$ are listed in Tab.~\ref{tab1}. As shown in Fig.~\ref{fig4}, the third-order coefficient $q_3$, arising from the octupole contribution, can be up to six times larger than the conventional AHE coefficient ($q_1$) at its maximum for (Ag$_{2}$Fe$_{5}$)$_8$. This is significantly greater than the octupole contribution ($q_3$) observed in the Fe$\mid$Ag single interface, where the maximum value is approximately 12\%. Morover, the different $q_3$ of (Ag$_{2}$Fe$_{5}$)$_{n}$ ($n=6,8,10$) exhibit a pronounced distinction under periodic modulation, which suggests that the magnitude of the octupole moment can be tunable. And, we have constructed an antiferromagnetic state by making the magnetic moment of adjacent \((\text{Ag}_{2}\text{Fe}_{5})\) units antiparallelly arranged, leading to a sign reversal of the octupole contribution. Additionally, the values of the higher-order coefficients are all smaller than that of the third-order coefficient $q_3$ and exhibit a decreasing trend. However, the ninth-order term remains non-negligible, suggesting that higher-order multipole moment also make a certain contribution, which should be studied in the future.



In conclusion, we introduce a novel approach for quantifying the octupole moment, which relies on angular-dependent anomalous Hall current and incorporates the discrete crystal symmetry. Unlike the conventional Berry curvature framework, our methodology enables the investigation of octupole contribution in non-periodic interfaces and surfaces. Using this method, we demonstrate octupole-dominated anomalous Hall effects in (Ag$_{2}$Fe$_{5}$)$_{n}$ multilayer systems, which are designed to construct quantum well for selecting $k$-points. Additionally, our results demonstrate that effective modulation of octupole moment strength can be achieved through interface engineering and magnetic ordering reconfiguration, laying a critical foundation for designing next-generation spintronic devices.

\begin{acknowledgments}
This work is financially supported by the National Key Research and Development Program of China (Grant Nos. 2023YFA1406600).
\end{acknowledgments}

\bibliography{ref.bib}

\end{document}